\documentclass[]{aa}
\usepackage{graphicx}
\usepackage{txfonts}
\usepackage[colorlinks]{hyperref}
\hypersetup{colorlinks=true,linkcolor=blue,citecolor=blue,filecolor=blue,urlcolor=blue,}
\usepackage{amsmath}
\usepackage{algorithm}
\usepackage{amsfonts}
\usepackage{multirow}
\usepackage{mathrsfs}
\usepackage{comment}
\usepackage{bm}
\usepackage{rotating}
\usepackage{color}
\usepackage{graphicx}
\usepackage{subfigure}
\usepackage{savesym}
\usepackage[flushleft]{threeparttable}
\savesymbol{tablenum}
\restoresymbol{SIX}{tablenum}

\newcommand{\at}[1]{{\textcolor{magenta}{\bf [AT: #1]}}} 
\usepackage{lineno}

\begin{document}

   \title{Exploring the AGN population in protoclusters: results from the TNG300 simulation and comparison with observations}

    \author{A. Traina \inst{1}, F. Vito \inst{1}, A. Pillepich \inst{2}, A. Kapahtia \inst{2}, O. Cucciati \inst{1}, R. Gilli \inst{1}, F. Arrigoni-Battaia \inst{3}, C. Vignali \inst{4,1}, M. N. Isla Llave \inst{1,4}, V. L. Cavicchi \inst{1,4}}
    
   \institute{Istituto Nazionale di Astrofisica (INAF) - Osservatorio di Astrofisica e Scienza dello Spazio (OAS), via Gobetti 93/3, I-40129 Bologna, Italy
        \and 
        Max-Planck-Institut für Astronomie, Königstuhl 17, D-69117 Heidelberg, Germany
        \and 
        Max-Planck-Institut für Astrophysik, Karl-Schwarzschild-Str. 1, D-85748 Garching, Germany 
        \and
        Dipartimento di Fisica e Astronomia (DIFA), Universit\`a di Bologna, via Gobetti 93/2, I-40129 Bologna, Italy
        }

   \date{Received ??; accepted ??}

 
  \abstract
    {Present-day galaxy clusters evolved from high-redshift overdensities ($z > 2$). Galaxy and supermassive black hole (SMBH) formation and evolution in these environments are expected to be accelerated by environmental effects, such as enhanced merger rates and large gas reservoirs. However, current observational studies aimed at characterizing this enhancement are limited by heterogeneous selection criteria for both protoclusters and AGN, as well as by the difficulty in confirming them spectroscopical. As a consequence, a robust and unbiased characterization of galaxy growth and AGN activity in dense environments is still lacking.}
    {In this work, we investigate the physical properties and cosmic evolution of galaxies and AGN in a sample of 280 protoclusters identified in the TNG300 simulation, selected to end up in $z = 0$ clusters with $M_{200, \rm c} > 10^{14} \, {\rm M_{\odot}}$. Our goal is to provide the first statistical view of AGN activity enhancement in a uniformly defined sample of overdense environments, and to identify the physical mechanisms driving it.}
    {We identify protoclusters as the progenitors of present-day galaxy clusters through merger-tree reconstruction and compare their galaxy and AGN populations with a control sample of field galaxies across the redshift range $0 \leq z \leq 6$. We investigate galaxy and SMBH demographics, AGN fractions, bolometric luminosity functions, and the SMBH accretion rate density, consistently applying homogeneous selection criteria in all environments.}
    {{We find that TNG300 protoclusters host systematically more massive galaxies and SMBHs than the field by up to $1 \, {\rm dex}$ at all redshifts, with signatures of accelerated galaxy evolution already visible at $z \sim 3-4$. The AGN fraction increases with stellar mass in both environments and, at fixed host-galaxy stellar mass, is broadly consistent between protoclusters and field galaxies, indicating no strong environmental triggering of SMBH accretion. However, when analysed as a function of redshift, protoclusters exhibit a significant enhancement of AGN activity (by a factor $> 2$), particularly at high luminosities and early cosmic times. We show that this enhancement primarily arises from differences in the stellar-mass distributions of galaxies in overdense regions, where massive systems assemble earlier than in the field. Consistently, protoclusters dominate the bright end of the AGN bolometric luminosity function and contribute up to $\sim 50\%$ of the total SMBH accretion rate density at $z \sim 6$.}}  
    {}


   \titlerunning{AGN in protoclusters simulations}
   \authorrunning{A. Traina et al.}
   \maketitle   
%

\section{Introduction}\label{sec:intro}
The formation and evolution of galaxies are strongly influenced by their surrounding environment. The most massive gravitationally bound baryonic structures in the Universe, i.e. galaxy clusters, host the most evolved galaxy populations \citep[e.g.,][]{alberts_noble2022}, suggesting that their progenitors --- galaxy protoclusters --- underwent accelerated evolutionary processes \citep[see e.g.,][]{overzier2016, chaing2017}. Over the past decade, observations have revealed the presence of extended Ly$\alpha$ nebulae in the cores of protoclusters \citep[e.g.,][]{geach2009, cantalupo2014, hennawi2015, arrigoni2018, arrigoni2019, umehata2019}, as well as significant overdensities of dust-rich sub-millimeter galaxies \citep[e.g.,][]{dannerbauer2014, oteo2018, arrigoni2023, pensabene2024,herwig2025}. Together, these observations indicate that large reservoirs of gas are present in protoclusters and within their member galaxies. In addition, merger rates in protoclusters have been found to be higher than in the field \citep{liu2023, giddings2026}. The combination of abundant gas supply and enhanced merger activity is expected to promote gas inflows toward the central regions of galaxies, thereby fueling supermassive black hole (SMBH) growth and triggering active galactic nucleus (AGN) activity \citep[e.g.,][]{hennawi2015, marchesi2023, vito2024}. 

\par Currently, only a limited number of studies have investigated the AGN population in overdense environments, with most works focusing on X-ray selected samples \citep[e.g.,][]{vito2020, tozzi2022a, vito2024, traina2025, travascio2025,IslaLlave2026}. These studies generally report an enhancement of AGN activity in protoclusters, as inferred from both the X-ray luminosity function (XLF) and the AGN fraction. \citet{traina2025} compared the overdensity factors of AGN and sub-millimeter galaxies (SMGs) across a sample of protoclusters, finding that AGN are systematically more overdense by up to a factor of $\sim 100$ than the parent population of SMGs. This result further supports the idea that SMBH growth may be particularly efficient in overdense environments. However, current observational studies are affected by significant inhomogeneities in the selection of galaxies, AGN, and protoclusters. These differences make it challenging to disentangle genuine environmental effects from selection biases, and therefore to robustly interpret the origin of the observed AGN enhancement.

\par A significant step forward in this direction has been recently made by \citet{IslaLlave2026}, who constructed, for the first time, a uniformly selected catalog of protoclusters, combining ALMA-observed SMGs and {\it Chandra}-detected AGN, and comparing their properties with well-defined control samples. Their analysis reveals an enhancement of AGN activity in protoclusters that appears to be independent of the physical properties of both galaxies and AGN, suggesting a genuine environmental effect. Similarly, \citet{shah2024}, by selecting a sample of spectroscopically confirmed massive protoclusters, found an enhancement of star formation activity in the overdense environments with respect ot the field. Nevertheless, the size of current observational samples remains limited, and it is still challenging to robustly isolate the impact of environment on SMBH accretion and activity from statistical uncertainties and residual selection biases.

\par A unique opportunity to investigate the impact of environment on galaxies and SMBHs is provided by cosmological hydrodynamical simulations of galaxies, where physical properties and evolutionary histories can be consistently tracked across cosmic time. Several studies have explored the evolution of simulated protoclusters, primarily focusing on their star formation properties \citep[e.g.,][]{bassini2020, lim2021, remus2023}, highlighting tensions between simulations and observations in reproducing the most extreme star-forming overdensities \citep[see also][for a comprehensive comparison of predicted and observed star formation rate functions]{gruppioni2015sam, katsianis2017eagle, katsianis2021sfrd, traina2026}. More recently, \citet{baxter2025} presented a systematic study of the identification and demographic properties of protoclusters using the TNG-Cluster simulation \citep{nelson2024}. However, a detailed and statistically robust investigation of SMBH growth and AGN activity in protoclusters, and their comparison with the field across cosmic time, is still lacking.

\par In this work, we use the publicly available IllustrisTNG suite of cosmological hydrodynamical simulations \citep[][]{nelson2019}\footnote{\href{https://www.tng-project.org/}{https://www.tng-project.org/}} to characterize the physical and statistical properties of SMBHs in protoclusters, and to investigate the origin of the reported AGN enhancement in overdense environments. We perform a systematic comparison between protocluster and field populations across cosmic time, analyzing galaxy and SMBH demography, AGN fractions, bolometric luminosity functions, and the SMBH accretion rate density. This approach allows us to assess the relative role of environment and host-galaxy properties in shaping SMBH growth and activity {in a well-defined volume-limited sample}.

\par The paper is organized as follows. In Section \ref{sec:simulations}, we describe the simulation used in this work and how we select clusters and protoclusters. Results on the SMBH demography and AGN enhancement are presented in Section \ref{sec:results}. Finally, we summarize our conclusions in Section \ref{sec:conclusion}. In this work, we assume a \cite{chabrier2003imf} stellar initial mass function (IMF) and adopt a $\Lambda$CDM cosmology with $H_{0} = 67.8$ $\rm km$ $\rm s^{-1}$ $\rm Mpc^{-1}$, $\Omega_{\rm m} = 0.3$, and $\Omega_{\Lambda} = 0.7$ \citep[][]{planck2016}.

\section{The simulation: TNG300 of IllustrisTNG}\label{sec:simulations}
The IllustrisTNG project \citep[see][for a complete description of the project]{marinacci2018, naiman2018, nelson2018,nelson2019tng50,nelson2019, pillepich2018illustris, pillepich2019tng50, springel2018} is a suite of large-volume cosmological magneto-hydrodynamical simulations carried out with the \texttt{AREPO} moving-mesh simulation code \citep[][]{springel2010, weinberger2020}. In the simulation, the cosmological parameters from the \citet{planck2016} are adopted. Dark matter (DM) halos are identified via the Friends-of-Friends algorithm \citep[][]{davis1985DM}, while the gravitationally bound subhalos are identified using the \texttt{SUBFIND} code \citep[][]{springel2001}. Across simulation snapshots, progenitors and descendants are linked through the \texttt{SubLink} merger tree algorithm \citep[][]{rodriguez-gomez2015}. The physical models implemented in the simulation include gas radiative processes, star formation \citep[modeled following][]{springel2003}, adopting a Schmidt--Kennicutt relation \citep{kennicutt1998sfr} and a \citet{chabrier2003imf} stellar IMF, evolution of stellar populations, outflows from supernovae (SNe) and SMBH physics \citep[][]{weinberger2017, pillepich2018illustris}.

\par In this work, we use the publicly available data from the TNG300 simulation of the IllustrisTNG project. TNG300 is the largest realization of the main IllustrisTNG simulation suite, which also includes TNG50 and TNG100. The simulated volume spans \(V = 302.6^{3}\,{\rm cMpc^{3}}\), enabling the formation of massive structures such as galaxy groups and clusters in statistically significant numbers. Among the three available realizations (TNG300-1, TNG300-2, and TNG300-3), we perform our analysis on the highest-resolution run, TNG300-1 (hereafter TNG300), which contains $(2500)^3$ dark matter and gas resolution elements. The baryonic and dark matter mass resolutions are $m_{\rm b}=1.1\times10^{7}\,{\rm M_{\odot}}$ and $m_{\rm DM}=5.9\times10^{7}\,{\rm M_{\odot}}$, respectively. Finally, the simulation outputs are available in 100 snapshots spanning $0 \leq z \leq 20$, with time intervals ranging from approximately 50 to 200 Myr \citep{nelson2019}.
A key advantage of TNG300 for this work is its large cosmological volume, which provides a statistically significant and volume-limited sample of present-day galaxy clusters together with their progenitors identified through merger-tree reconstruction. This allows us to follow the evolution of a homogeneous population of structures from high redshift to the present day, while minimizing selection effects \citep[e.g.,][]{pillepich2018results, donnari2021a}. Although dedicated cluster simulations such as TNG-Cluster \citep{nelson2024} provide access to larger samples of massive clusters and extend to higher halo masses, their zoom-in nature introduces a more complex selection function that is not strictly volume-limited. For this reason, TNG300 offers a particularly clean framework for studying environmental effects on galaxy and SMBH evolution.

\subsection{SMBH modeling and AGN growth and feedback}
A comprehensive and detailed description of the AGN modeling can be found in \citet{weinberger2017}. Here we summarize the main assumptions on the accretion onto the SMBHs and the subsequent feedback. SMBHs are treated as a type of particle in the simulation: a new SMBH is seeded inside a massive halo ($M_{\rm halo} > 7.4 \times 10^{10}\,{\rm M_{\odot}}$) with $M_{\rm BH} = M_{\rm seed} = 1.2 \times 10^6 \, {\rm M_{\odot}}$, whenever the friend-of-friend algorithm finds a massive halo without a BH. 

The growth of SMBHs occurs through gas accretion and SMBH mergers. Gas accretion is computed using a Bondi--Hoyle--Lyttleton prescription and is capped at the Eddington limit. While the accretion prescription remains the same at all accretion rates, the AGN feedback model operates in two distinct states, commonly referred to as the high- and low-accretion states \citep[see e.g.,][]{sijacki2007}. The transition between the two states is determined by the Eddington ratio,
\begin{equation}
    \frac{\dot{M}_{\rm Bondi}}{\dot{M}_{\rm Edd}} = \chi,
\end{equation}
where $\chi$ is a threshold parameter. SMBHs above this threshold enter the high-accretion state, while those below it operate in the low-accretion state. Different feedback channels and efficiencies are associated with the two states, with the injected energy depending on the SMBH accretion rate and radiative efficiency ($\epsilon_{\rm r}\sim 0.1 - 0.2$). In the high-state, gas is subjected to thermal, isotropic heating. In the lower state, instead, episodic kinetic feedback act by removing gas nearby the SMBHs.

Despite some models assume a fixed threshold \citep[e.g.,][]{sijacki2007,sijacki2015bhard, vogelsberger2013, vogelsberger2014}, the AGN model in IllustrisTNG allows it to vary with the SMBH mass:
\begin{equation}
    \chi =  {\rm min} \left[ \chi_{0} \left(\frac{M_{\rm BH}}{10^8 \, {\rm M_{\odot}}} \right)^{\beta}, 0.1 \right]
\end{equation}
where $\chi_{0}$ and $\beta$ are model free parameters, with fiducial values $\chi_{0} = 0.002$ and $\beta=2$.

The comparison between the AGN LF in TNG300 and observations is presented in \citet{habouzit2022} and \citet{Kapahtia2026}.

\subsection{Selection of galaxies in (proto)clusters and field environment}\label{subsec:clusters_selection}
\begin{figure}[]
\centering
\includegraphics[width=.5\textwidth]{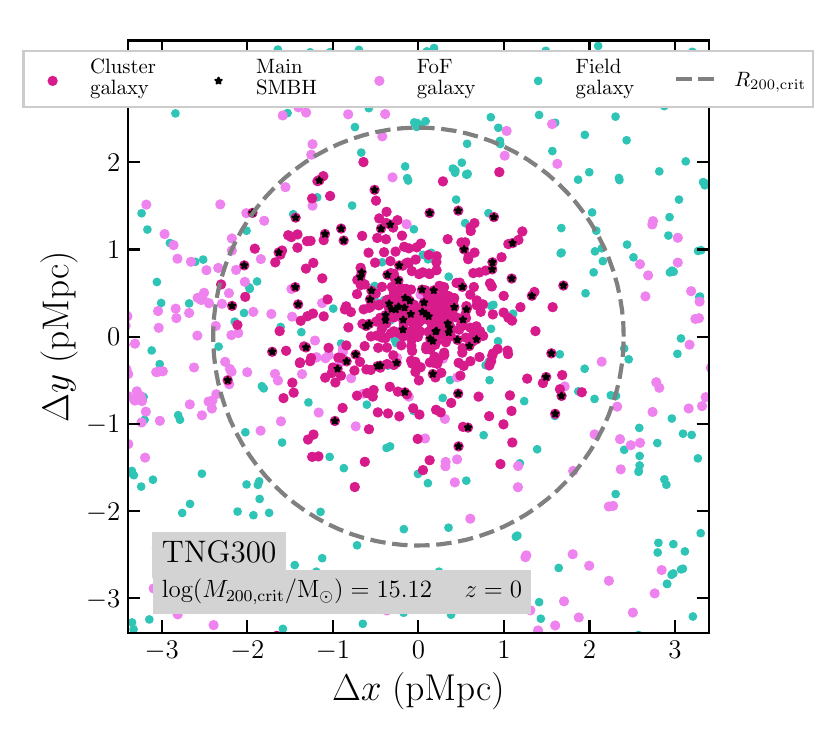}
 \vspace{-0.5cm}
 \caption{Example of a galaxy cluster a $z = 0$ in the TNG300 simulation {($x$ and $y$ sizes are $~7 \,{\rm pMpc}$, while there is no cut in the $z$ axis)}. The grey dashed line is the circle corresponding to $R_{\rm 200, crit}$. Red and pink markers are the galaxies belonging to the massive halo ($M_{\rm 200, crit} = 1.32 \times 10^{15} \, {\rm M_{\odot}}$) inside and outside the critical radius, respectively. Black stars are the main SMBHs of the cluster galaxies. Finally, the green circles are the galaxies belonging to the field.}
 \label{fig:cluster_definition}
\end{figure}

\begin{figure*}[]
\centering
\includegraphics[width=1.\textwidth]{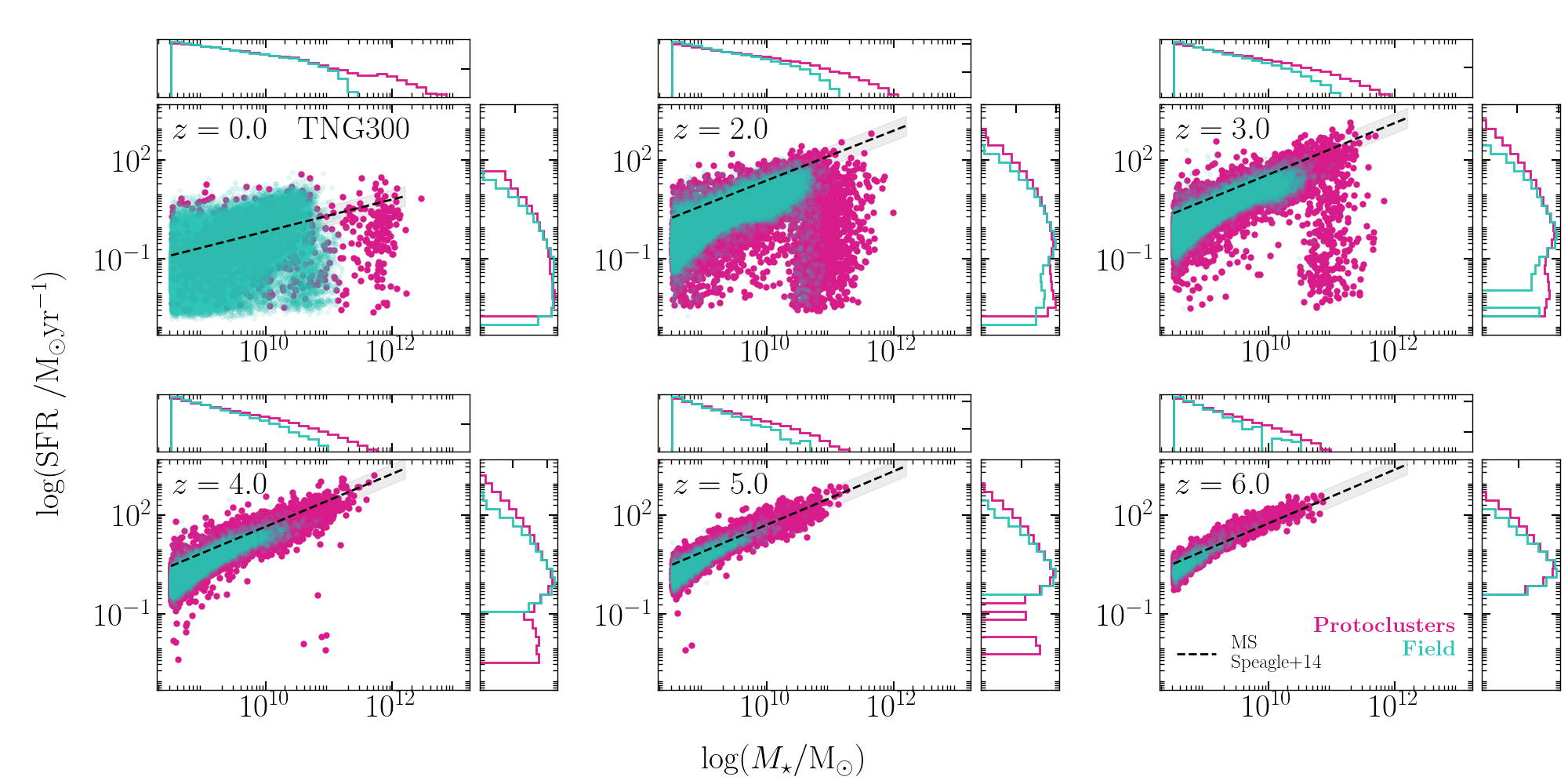}
 \caption{SFR - $M_{\star}$ distributions at different redshifts of galaxies in the TNG300 simulation. Green points are field galaxies, while red markers denote galaxies in clusters at $z = 0$ and in their “progenitor” protoclusters at $z > 0$. The $x$-axis and $y$-axis histograms are the normalized distributions of the individual quantities for protoclusters and field. Protoclusters galaxies can reach larger masses and are more affected by AGN quenching in the high masses. Black dashed line is the main sequence of galaxies by \citet{speagle2014MS} for comparison.}
 \label{fig:SFR_Mstar}
\end{figure*}

In order to investigate possible environmental effects on the growth of SMBHs and AGN activity, we conduct our analysis on the progenitors of local (i.e., at $z = 0$) galaxy clusters. Such clusters are selected in the snapshot $\#99$ (corresponding to $z = 0$) as those halos with $M_{\rm 200,crit} > 10^{14}\, {\rm M_{\odot}}$, where $M_{\rm 200,crit}$ is the mass enclosed in a sphere with radius $R_{\rm 200,crit}$, corresponding to a mean density $\bar{\rho} = 200\, \rho_{\rm crit}$. A different definition of radius (e.g., $R_{500, {\rm crit}}$) would change the cluster sample, reducing it to the most massive structures only. The final sample of clusters identified through our definition is 280, with most of them having $10^{14}<M_{\rm 200,crit} < 10^{14.5}\,{\rm M_{\odot}}$ and just a few of them reaching larger masses \citep[see][]{nelson2024}.
Following \citet{baxter2025}, we define cluster galaxies as all \texttt{SUBFIND} subhaloes with stellar mass $M_{\star}>10^{8.5}\,{\rm M_{\odot}}$ that belong to the cluster FoF halo and are located within $R_{\rm 200,crit}$ of the cluster centre. Galaxies associated with the same FoF halo but residing outside $R_{\rm 200,crit}$ are not considered cluster members and are excluded from all subsequent analysis. Figure \ref{fig:cluster_definition} shows an example of a galaxy cluster at $z = 0$ with ${\rm log}(M_{\rm 200,crit}/{\rm M_{\odot}}) = 15.12$. Red and pink points mark FoF galaxies inside and outside the sphere with radius $R = R_{\rm 200,crit}$, respectively. Thus, only red points are considered as cluster members.
Following the identification of the $z=0$ cluster sample, protoclusters are defined as the progenitors of these systems at higher redshifts. Specifically, we use the \texttt{SubLink} merger trees to trace all cluster member galaxies back in time and define a protocluster as the ensemble of galaxies that will eventually become members of the same $z=0$ galaxy cluster.

Since our main goal is to compare AGN in overdense environments with those in isolated environments, we define the field as the collection of galaxies that do not belong to halos that are either galaxy groups ($10^{13} < M_{\rm 200, crit} < 10^{14}\,{\rm M_{\odot}}$) or clusters at $z = 0$, and their progenitors (at $z > 0$). Field galaxies are shown as green points in Figure \ref{fig:cluster_definition}.

We note that our definition of protoclusters is more restrictive than the one commonly adopted in observational studies. Observationally, protoclusters are often identified as overdense regions at high redshift using galaxy or AGN tracers, without direct knowledge of their future evolution. In contrast, our protocluster sample is selected through merger-tree reconstruction and therefore consists exclusively of high-redshift overdensities that are guaranteed to evolve into present-day galaxy clusters. As discussed by \citet{baxter2025}, not all high-redshift overdensities satisfy this condition. Our approach therefore provides a cleaner and more physically motivated sample for studying the environmental dependence of galaxy and SMBH evolution.

\subsection{Selection of SMBHs and AGN}\label{subsec:BH_selection}
Throughout this work, we will refer the analysis to the main SMBH of a galaxy. This is defined as the most massive SMBH residing in the galaxy (similar results can be obtained if using the SMBH nearest to the galaxy center). SMBHs with a bolometric luminosity above a certain threshold (we adopt different values in the following analysis) are classified as AGN. The black stars in Figure \ref{fig:cluster_definition} indicates the main SMBHs of cluster galaxies. 
\par We compute the AGN luminosity (which is not a direct output of the simulation) following the same approach of \citet{habouzit2022}, based on the model by \citet{churazov2005}. For radiatively efficient SMBHs (i.e., $f_{\rm edd} = {\dot M_{\rm BH}} / {\dot M_{\rm edd}} > 0.1$):
\begin{equation}
    L_{\rm BOL} = \frac{\epsilon_{\rm r}}{1 - \epsilon_{\rm r}} {\dot M_{\rm BH}} c^2 \, ,
\end{equation}
where ${\dot M_{\rm BH}}$ is the SMBH growth rate.
If $f_{\rm edd}<0.1$, the SMBH is radiatively inefficient and the bolometric luminosity is computed as:
\begin{equation}
    L_{\rm BOL} = 10f_{\rm edd} \epsilon_{\rm r} {\dot M_{\rm BH}} c^2,
\end{equation}
\\
with $\epsilon_{\rm r} = 0.2$.


\section{Results}\label{sec:results}

Previous studies based on the IllustrisTNG simulations have
characterized galaxy populations \citep[e.g.][]{pillepich2018results, pillepich2019tng50}, SMBH populations \citep{habouzit2022}, and the impact
of SMBH feedback on both galaxies \citep{nelson2018,donnari2019, donnari2021a, donnari2021b, Kurinchi-Vendhan2023} and AGN \citep{Kurinchi-Vendhan2025,Kapahtia2026} at various cosmic epochs and cosmological environments. Here, we revisit these findings and extend them by explicitly
contrasting protocluster and field galaxies, at $z=0 - 6$.

\subsection{Galaxies and SMBHs demography: protoclusters vs field}\label{subsec:AGN_gal_demo}

\begin{figure*}[]
\centering
\includegraphics[width=1.\textwidth]{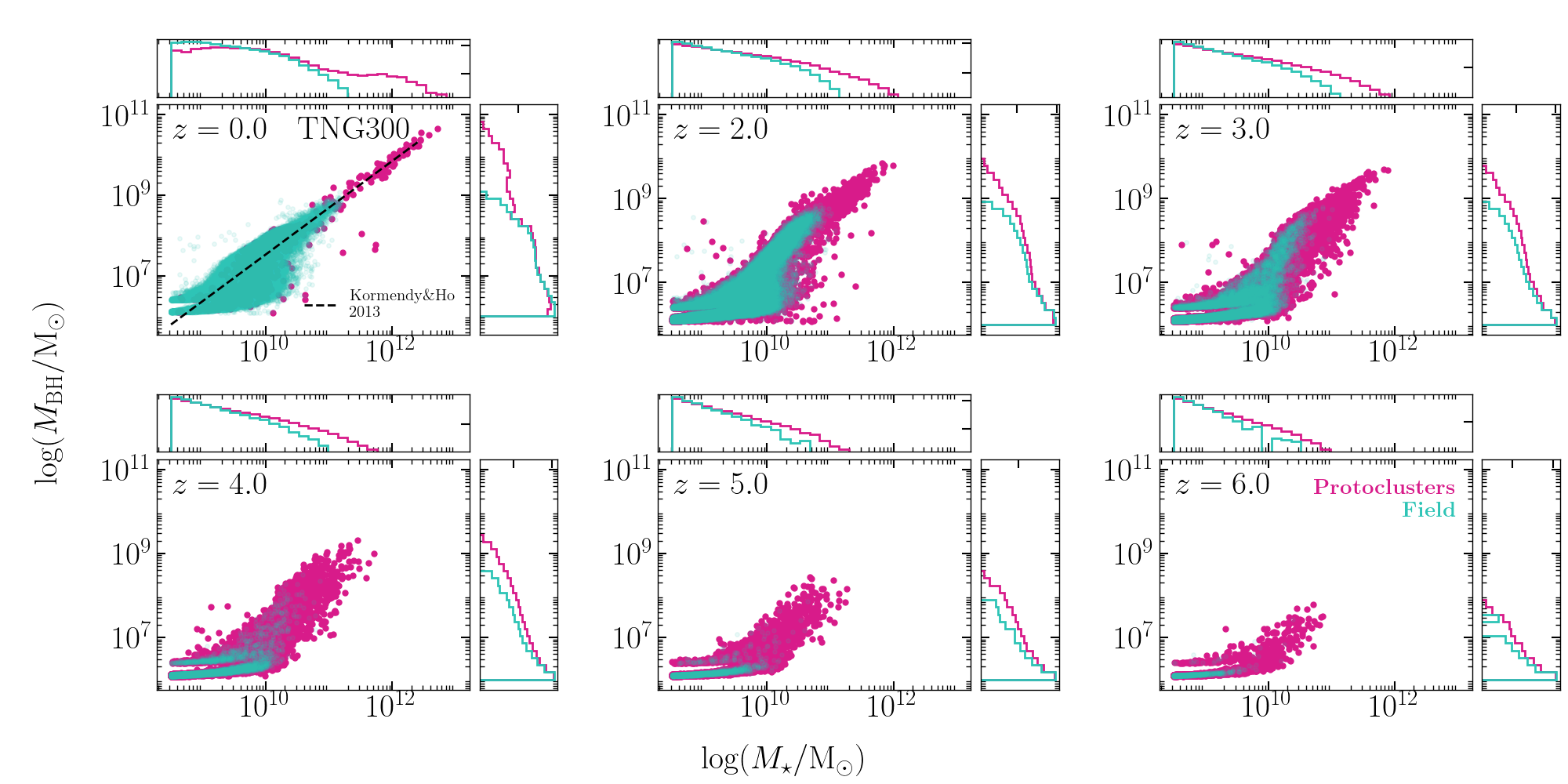}
 \caption{Same as Figure \ref{fig:SFR_Mstar} but for $M_{\rm BH} \,-\, M_{\star}$, i.e., the relation between the SMBHs and galaxies masses in the TNG300 simulation. More massive SMBHs are typically hosted in more massive galaxies, at each redshifts. as a comparison, we overplot the relation by \citet{kormendy2013coev} at $z= 0$.}
 \label{fig:MBH_Mstar}
\end{figure*}
\begin{figure*}[]
\centering
\includegraphics[width=1.\textwidth]{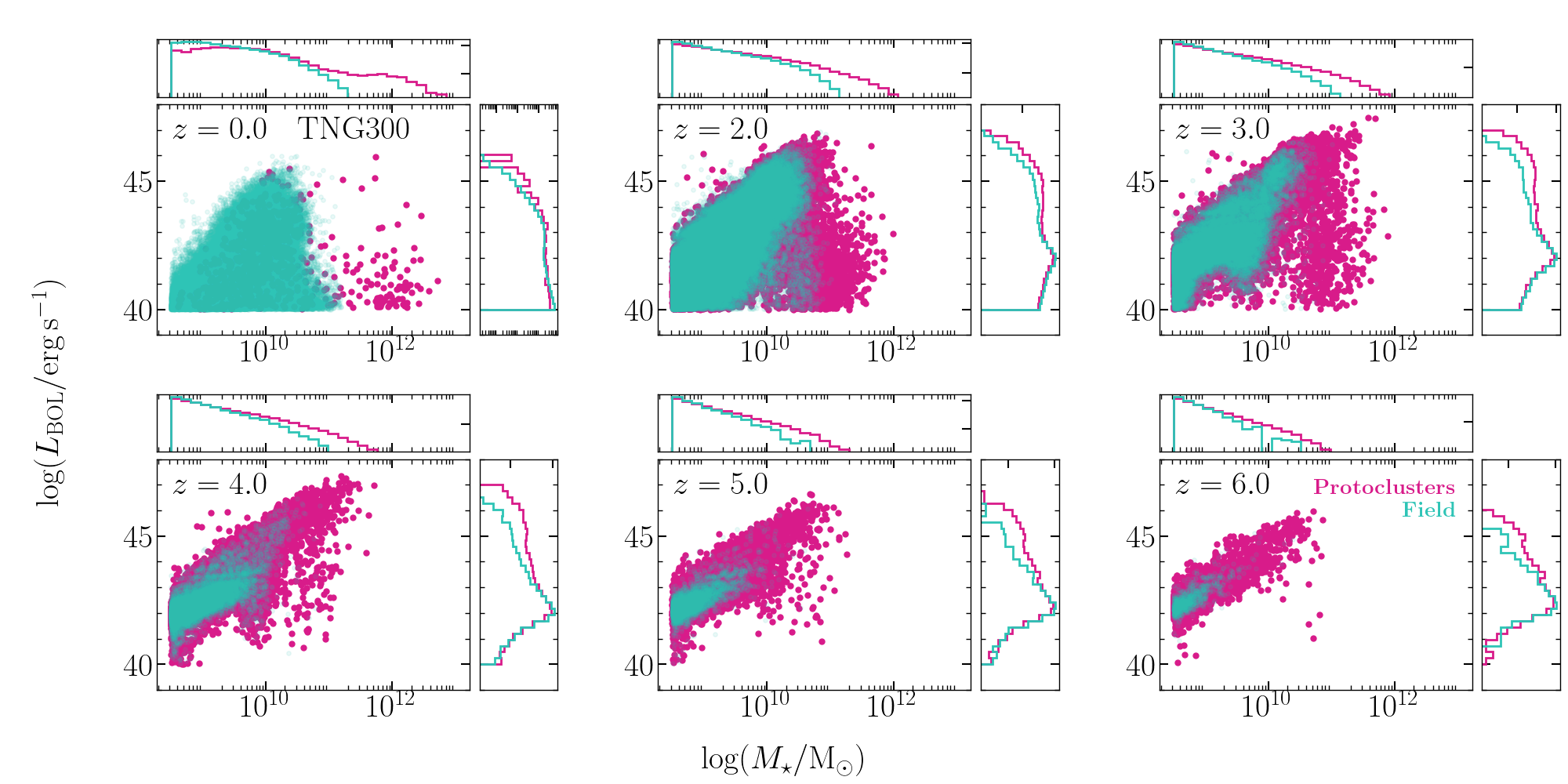}
 \caption{Same as Figure \ref{fig:SFR_Mstar} but for $L_{\rm BOL} \,-\, M_{\star}$, i.e., the distributions of SMBHs bolometric luminosity and galaxy stellar mass. More massive galaxies host brighter AGN, especially at high redshifts, where quenching is not yet in place.}
 \label{fig:Lbol_Mstar}
\end{figure*}
\begin{figure*}[]
\centering
\includegraphics[width=1.\textwidth]{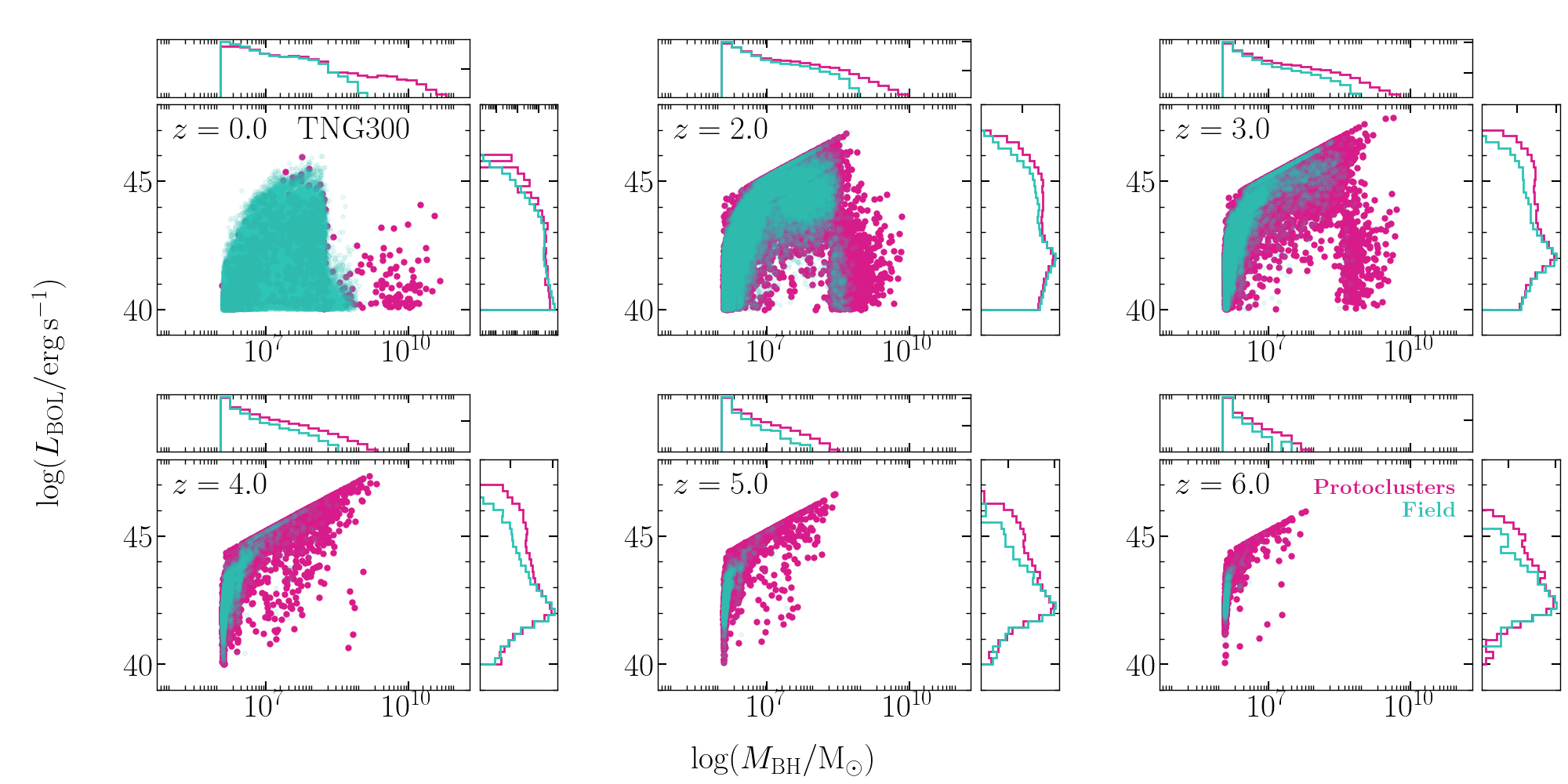}
 \caption{Same as Figure \ref{fig:SFR_Mstar} but for $L_{\rm BOL} \,-\, M_{\rm BH}$, i.e., the SMBH mass-luminosity relation. At high redshifts, more massive SMBHs are also brighter, while going towards lower redshifts quenching starts taking place and the most massive SMBHs are not the brightest.}
 \label{fig:Lbol_MBH}
\end{figure*}
We compare a set of physical properties for AGN and galaxies in different environments, at $0 \leq z \leq 6$. All the results shown in Figures \ref{fig:SFR_Mstar} to \ref{fig:Lbol_MBH} are obtained by assuming a minimum bolometric luminosity $L_{\rm BOL} > 10^{40} \, {\rm erg \, s^{-1}}$ (for the SMBH properties) and minimum stellar mass $M_{\star} > 10^{8.5} \, {\rm M_{\odot}}$. 

\subsubsection{Galaxies properties}\label{subsubsec:Gal_prop}
Figure \ref{fig:SFR_Mstar} shows the SFR vs $M_{\star}$ plane of galaxies at different redshifts. At $z = 6$, a well defined MS is already in place: the bulk of galaxies is at $M_{\star} < 10^{10} \, {\rm M_{\odot}}$, with the most massive ones reaching $M_{\star} \sim 10^{11} \, {\rm M_{\odot}}$, while their SFRs are in the range $\sim 1 \, -\, 100 \, {\rm M_{\odot}\, yr^{-1}}$. Towards lower redshifts, stellar masses increase, reaching up to $\sim 10^{12} \, {\rm M_{\odot}}$ at $z = 0$ (likely being BCG-like systems), while the SFRs distribution broadens down to $\sim 10^{-2} \, {\rm M_{\odot}\, yr^{-1}}$, tracing the transition from the star-forming to the quenched regime where galaxies start to populate the region below the main sequence. This trend reflects gas consumption and the progressive quenching of star formation from early epochs to the local Universe, from larger to smaller mass galaxies, and confirms previous studies based on the IllustrisTNG simulation \citep[e.g.,][]{donnari2019, pillepich2019tng50, donnari2021a, donnari2021b}
\par Although a similar evolutionary trend is observed in both protocluster and field galaxies, significant differences emerge in the stellar mass distributions of the two samples. At all redshifts, protoclusters and field have similar stellar mass distributions until $M_{\star} \sim 10^{11} {\rm \, M_{\odot}}$, but more massive galaxies are present in protoclusters, with the difference becoming more pronounced at lower redshifts. This is expected, as protoclusters reside within the most massive dark matter haloes of the simulation.
Notably, protoclusters already host galaxies with low SFRs at $z \sim 3\,-\,4$, while similar systems appear in the field only at lower redshifts. This suggests an earlier suppression of star formation in overdense environments \citep[see also][]{donnari2021a}.

\subsubsection{SMBHs properties}\label{subsubsec:BH_prop}
Similarly to stellar masses, SMBHs grow from high to low redshifts as a consequence of gas accretion and multiple merger events (see Figure \ref{fig:MBH_Mstar}). Low masses SMBHs ($M_{\rm BH} \sim 10^6 \,-\, 10^9 \,{\rm M_{\odot}}$) are found in both environments, whereas the most massive ones ($M_{\rm BH} > 10^9 \,{\rm M_{\odot}}$) are exclusively hosted in protoclusters. This result is related to earlier gas supply and higher merger rate in overdensities (see Appendix \ref{app:gas_content}). The earlier seeding of SMBHs, due to an earlier growth of galaxies in protoclusters, may also be an important factor in explaining the larger SMBH masses in overdense environments at fixed cosmic epoch. Due to SMBHs mergers, a double tail feature is visible at low SMBH masses, where the BHs are seeded.
\par The SMBH bolometric luminosity, shown in Figure \ref{fig:Lbol_Mstar}, traces the accretion activity onto SMBHs. At $4 \leq z \leq 6$, higher luminosities are preferentially associated with more massive galaxies, while at lower redshifts the overall activity declines, following a trend similar to that observed for the SFRs, likely driven by gas depletion. Interestingly, in protoclusters highly accreting SMBHs with large $L_{\rm BOL}$ ($> 10^{44} \, {\rm erg \, s^{-1}}$) are already in place at $z \sim 6$, whereas field SMBHs reach comparable luminosities only at $z \sim 3$, implying earlier and faster SMBH growth in protoclusters, as {a consequence of} the presence of massive SMBHs at high redshift, as shown in Figure \ref{fig:MBH_Mstar}. 
\par Figure \ref{fig:Lbol_MBH} shows the $L_{\rm BOL} \,-\, M_{\rm BH}$ relation. Overall, higher activity (i.e., larger $L_{\rm BOL}$) is associated with more massive SMBHs. An exception is represented by the low-mass SMBH tail, roughly corresponding to the seeding mass, where $L_{\rm BOL}$ spans over 4 orders of magnitude. At large $L_{\rm BOL}$ and $M_{\rm BH}$, a cap is visible in the distributions, due to the Eddington limit in the simulation. We can also see the effect of AGN quenching on the high-mass end, likely due to the feedback suppression of the gas accretion onto the SMBH. A more detailed discussion can be found in \citet{Kapahtia2026}.

\subsection{AGN fraction}
Several studies in the literature have reported an enhancement of AGN activity in protoclusters, typically quantified in terms of the AGN fraction in overdense environments \citep[e.g.,][]{tozzi2022a, vito2020, shah2024,vito2024, traina2025, travascio2025}. Motivated by the demographic trends discussed in the previous Section, where protoclusters are shown to host, on average, more massive and more luminous SMBHs than the field (except for the already quenched SMBHs, that have large masses but lower luminosities), we now investigate this enhanced SMBH activity from the perspective of the AGN fraction.

In the following analysis, the AGN fraction in the field and in protoclusters is defined in a consistent manner. At each redshift or galaxy stellar-mass bin, we compute the AGN fraction as the ratio between the number of SMBHs with bolometric luminosity above a fixed threshold, \(L_{\rm BOL} > L_{\rm AGN}\) (i.e. \(N_{\rm AGN}\), with \(L_{\rm AGN} = 10^{44}\) or \(10^{45}\,{\rm erg\,s^{-1}}\)), and the total number of galaxies (i.e., without restrictions on $L_{\rm BOL}$) in the same bin, \(N_{\rm GAL}\):
\begin{equation}
f_{\rm AGN} =
\left.
\frac{N_{\rm AGN}}{N_{\rm GAL}}
\right|_{\rm fixed\; z\; and\; M_{\star}} \, .
\end{equation}

Uncertainties for both environments are estimated as the 16th and 84th percentiles of the distribution obtained via bootstrap resampling of the AGN population.

\subsubsection{AGN fraction as a function of $M_{\star}$}
\begin{figure*}[]
\centering
{\includegraphics[width=0.49\textwidth]{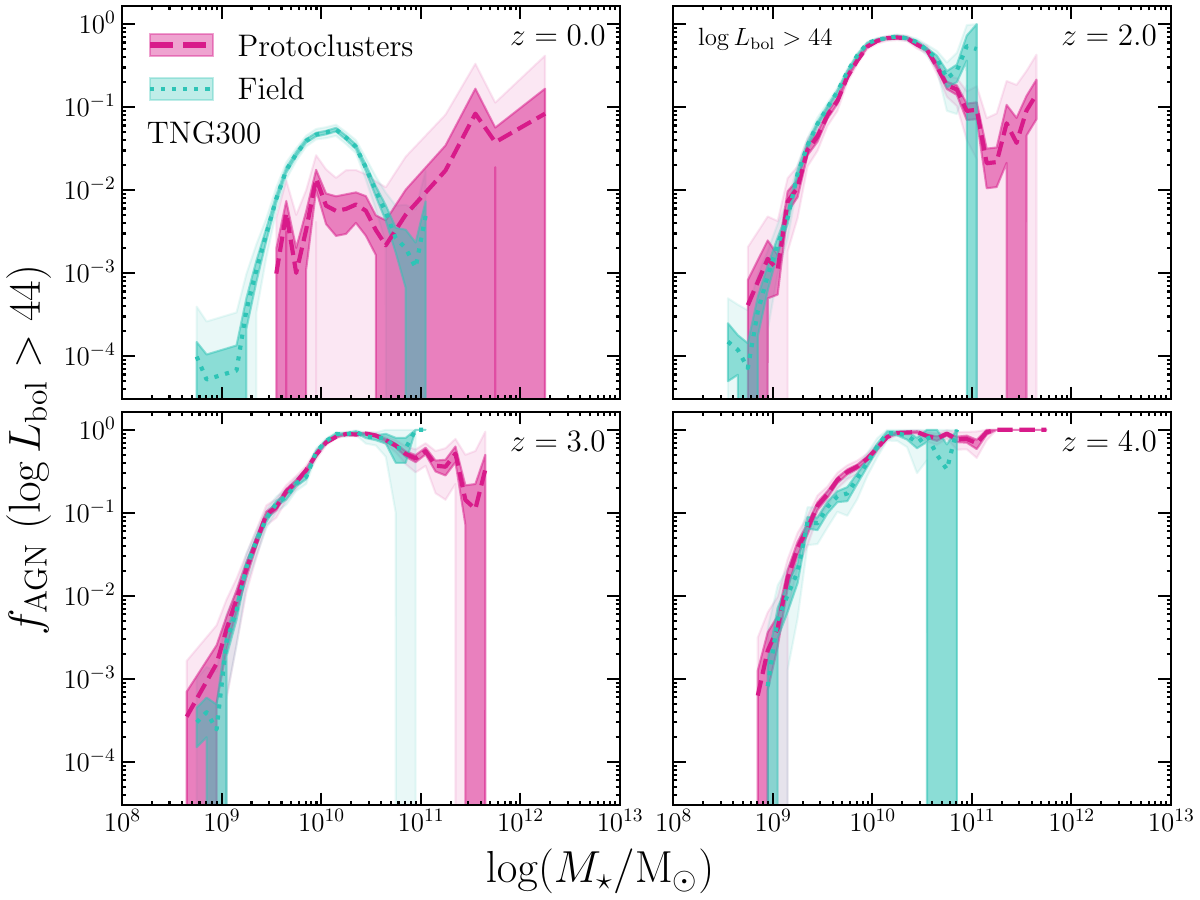}}
{\includegraphics[width=0.49\textwidth]{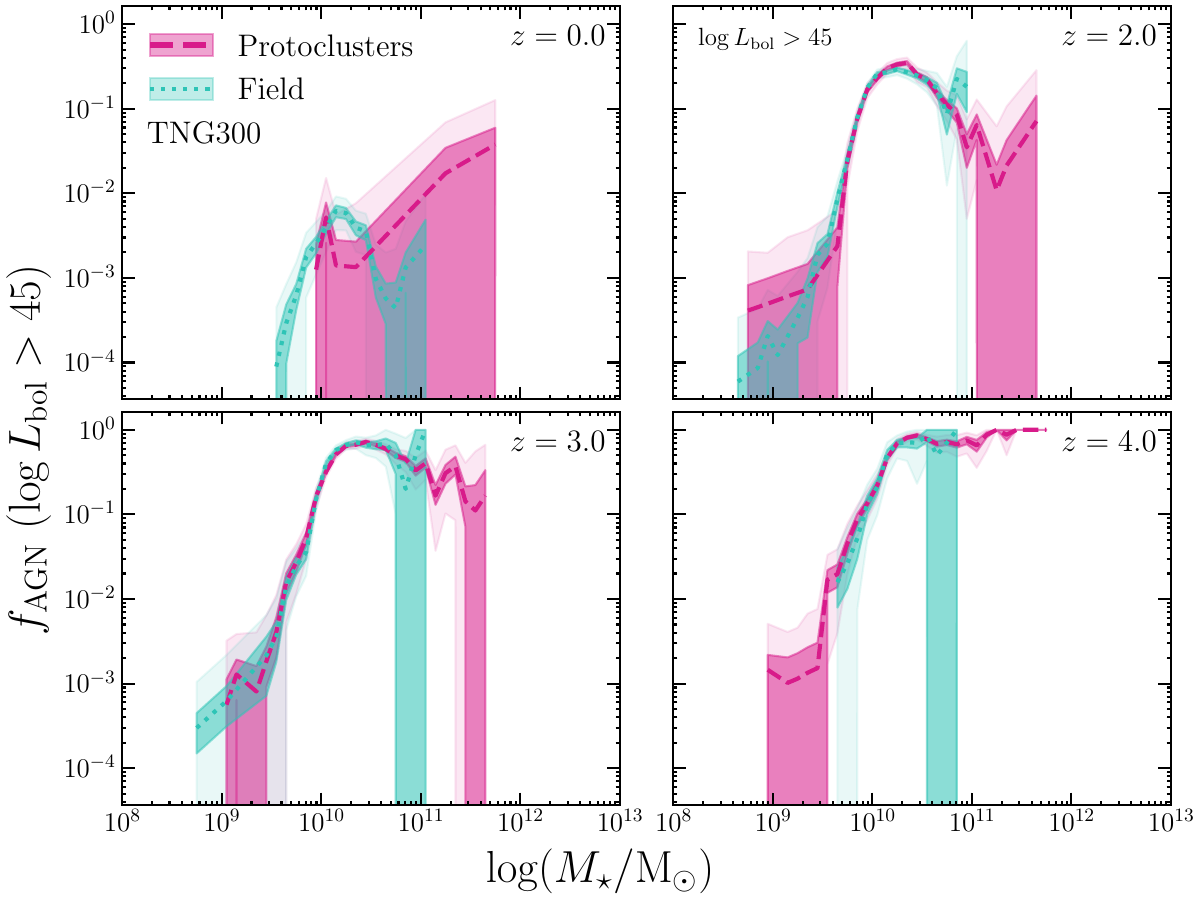}}
\caption{AGN fraction as a function of galaxy stellar mass in different redshift bins. The red and green curves are the median values for protoclusters and field, respectively, while the shaded areas are $1\sigma$ (darker) and $3\sigma$ (lighter) uncertainties. Left and right panel show results for different luminosity thresholds ($10^{44}$ and $10^{45}\, {\rm erg \, s^{-1}}$, respectively).} 
\label{fig:fagn_zbin}
\end{figure*}
Here we discuss the dependence of the AGN fraction on the galaxy stellar mass, at individual redshifts (\(z = 0,\,2,\,3,\,4\)), corresponding to the local galaxy cluster and protocluster phases. The results are shown in Figure \ref{fig:fagn_zbin}, with different \(L_{\rm AGN}\) thresholds adopted in the left and right panels ($L_{\rm BOL} > 10^{44}\,{\rm erg \, s^{-1}}$ and $L_{\rm BOL} > 10^{45}\,{\rm erg \, s^{-1}}$, respectively). Since the trends are very similar in the two panels, differing only by a mild normalization shift due to the adopted AGN luminosity threshold. In the following, we focus on the left, less conservative panel.

According to TNG300, at \(2 \leq z \leq 4\), the AGN fraction increases with stellar mass, consistently with previous observational studies \citep[e.g.,][]{magliocchetti2020}, from \(f_{\rm AGN} \sim 10^{-3}\) up to \(f_{\rm AGN} \sim 0.8\,-\,1\) going towards larger galaxy masses (\(M_{\star} \sim 10^{10}\,{\rm M_{\odot}}\)). At higher stellar masses, the AGN fraction flattens at \(z = 4\) and decreases at lower redshifts, with the decline becoming progressively steeper toward lower redshift. This behavior is found in both field and protocluster environments, although it is less pronounced in the field, where the stellar-mass range is more limited. However, this decrease of the AGN fraction is not typically found in observed clusters \citep[e.g.,][]{pimbblet2013, bitsakis2015, peluso2022} and can be ascribed to the AGN quenching at the high-masses and to massive galaxies hosting low-luminosity AGN \citep[see][]{Kurinchi-Vendhan2025, Kapahtia2026}.

At fixed stellar mass, the AGN fraction in protoclusters, \(f_{\rm AGN,\,PCs}\), is consistent with that measured in the field, \(f_{\rm AGN,\,Field}\), indicating no significant environmental enhancement of SMBH activity at fixed stellar mass. This consistency breaks down at \(z = 0\) \citep[see][for TNG]{Kurinchi-Vendhan2025}. In the field, the AGN fraction shows a trend similar to that observed at higher redshifts, with a peak at \(M_{\star} \sim 10^{10}\,{\rm M_{\odot}}\). In contrast, in galaxy clusters the AGN fraction increases almost monotonically with stellar mass, but with large uncertainties, and thus it is actually also consistent with a decline of $f_{\rm AGN}$ with stellar mass. This behavior is similar to that found for SFR in observations of clusters vs field \citep[e.g.,][]{vulcani2010, jian2018}.

\subsubsection{AGN fraction as a function of redshift}
\begin{figure*}[]
\centering
{\includegraphics[width=1\textwidth]{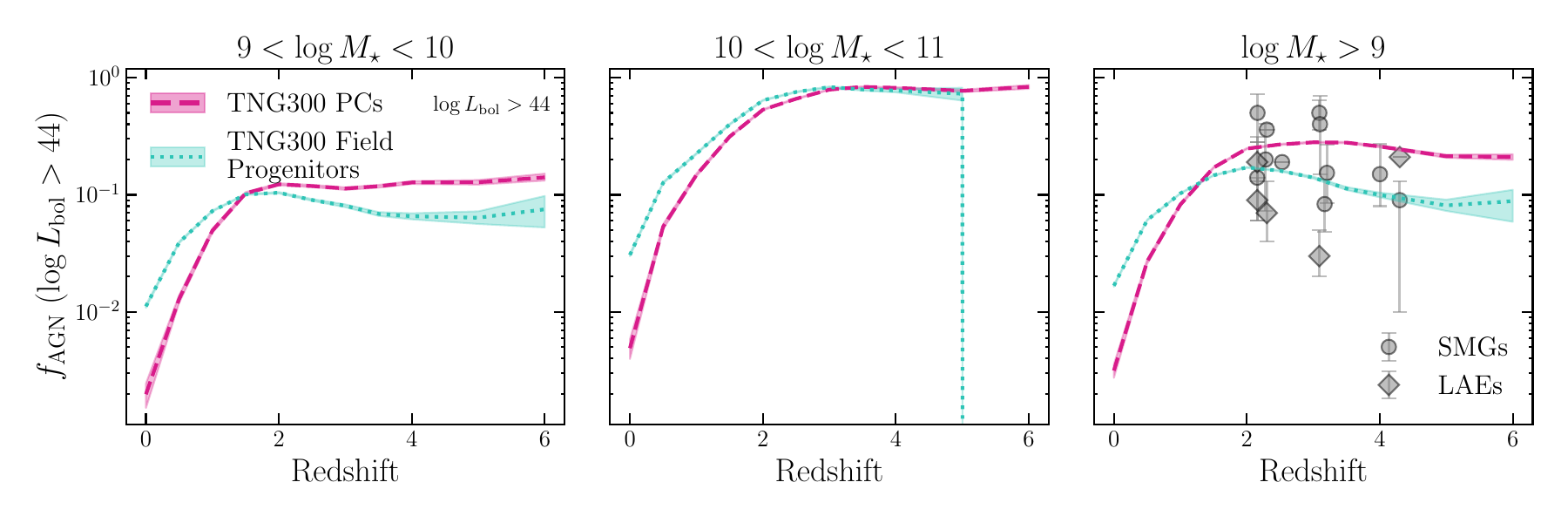}}
\vspace{-1cm}
{\includegraphics[width=1\textwidth]{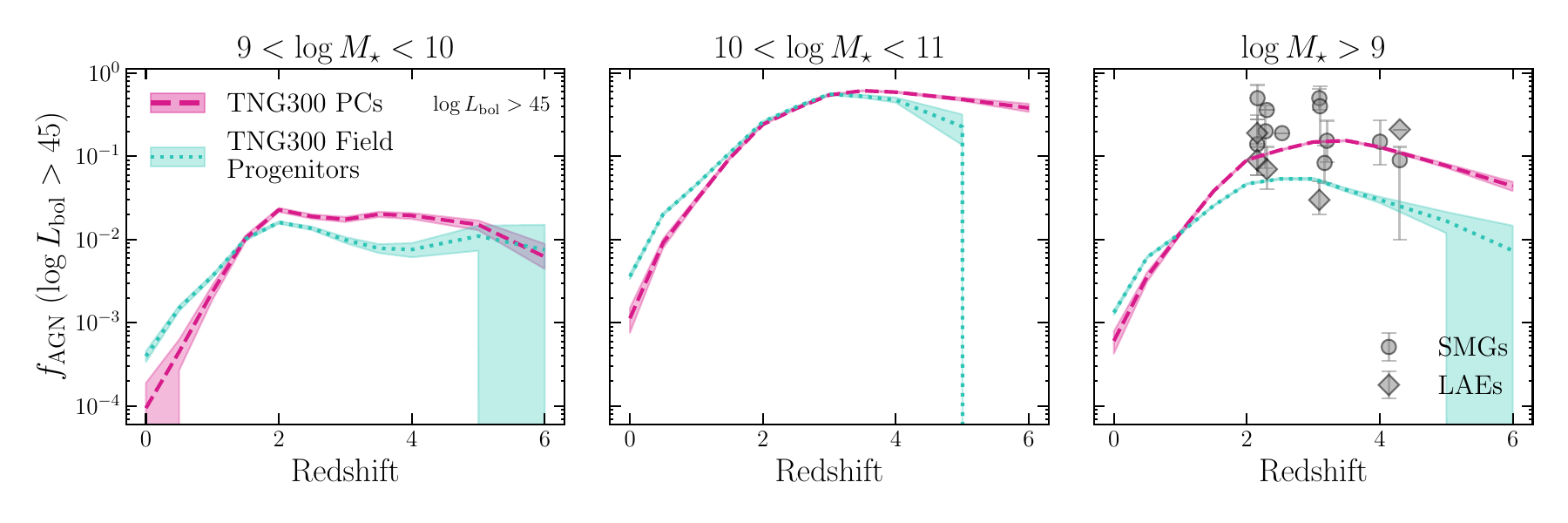}}
\caption{AGN fraction as a function of the redshift, in different galaxy stellar mass bins. Top and bottom panels show results obtained with different AGN luminosity thresholds ($10^{44}$ and $10^{45}\, {\rm erg \, s^{-1}}$, respectively). The red and green curves are the median values for protoclusters and field, while the shaded areas are $1\sigma$ uncertainties. Grey points are the AGN fractions in SMGs (circles) and LAEs (diamonds) derived for individual protoclusters by \cite{lehmer2009,digby-north2010,smail2014,chen2016,macuga2019,umehata2019,vito2020,polletta2021,tozzi2022a,monson2023,perez-martinez2023,vito2024,travascio2025, traina2025}.} 
\label{fig:fagn_mbin}
\end{figure*}

Similarly to the previous Section, we investigate the redshift evolution of the AGN fraction in bins of stellar mass. We divide the sample into two stellar-mass bins,
\(M_{\star} \in [10^{9},\,10^{10}]\,{\rm M_{\odot}}\) and
\(M_{\star} \in [10^{10},\,10^{11}]\,{\rm M_{\odot}}\),
and define an additional sample including all galaxies with \(M_{\star} > 10^{9}\,{\rm M_{\odot}}\).
The results are shown in Figure \ref{fig:fagn_mbin}, where the upper and lower panels correspond to different AGN luminosity thresholds.

In both the field and protocluster environments, the AGN fraction increases from \(z \sim 0\) up to \(z \sim 2\,-\,3\), and then flattens toward higher redshifts. At low redshift (\(z < 2\)), the AGN fraction in the field is higher than in protoclusters across all stellar-mass bins considered here. At higher redshifts, this trend is reversed, particularly at lower stellar masses.

In the mass bin \(M_{\star} \in [10^{9},\,10^{10}]\,{\rm M_{\odot}}\), the AGN fraction in protoclusters exceeds that in the field by a factor of \(\sim 2\) at \(2 < z < 6\), while it is higher in the field at lower redshifts. At higher stellar masses, the AGN fractions in the two environments are instead consistent within uncertainties over the same redshift range. When considering the full sample with \(M_{\star} > 10^{9}\,{\rm M_{\odot}}\), the discrepancy becomes more pronounced, primarily due to the inclusion of very massive galaxies (\(M_{\star} > 10^{11}\,{\rm M_{\odot}}\)), which are almost absent in the field. These trends are independent of the adopted AGN luminosity threshold. In this last bin we also show as a comparison the AGN fractions in observed sub-millimeter galaxies (SMGs) and Ly$\alpha$ emitters (LAEs), obtained for individual protoclusters. The comparison with observational measurements reveals a broad dispersion in the reported AGN fractions at all redshifts. This large scatter is expected, given the substantial heterogeneity among current observational studies. Different works adopt distinct luminosity thresholds, stellar-mass cuts (typical masses for these systems are above $10^{10}\, {\rm M_{\odot}}$), and protocluster identification methods, often probing different regions and evolutionary stages of overdense structures. Simulated protoclusters typically have an angular radius between $5^\prime$ and $10^\prime$ at $z \sim 3$, while observed protoclusters are extended from arcsecond up to arcminute scale \citep[see e.g.,][]{baxter2025}.
In addition, observational samples are still limited in size and are typically affected by incompleteness and cosmic variance. These differences make direct comparisons between individual studies intrinsically difficult and may partially account for the reported discrepancies in the level of AGN enhancement across the literature. In this context, cosmological simulations provide a complementary framework, where AGN and galaxy populations can be selected homogeneously and consistently across cosmic time and environments. The comparison between simulations and observations should therefore be interpreted primarily in terms of broad evolutionary trends and relative environmental differences, rather than aiming for an exact reproduction of individual observational measurements.

\begin{figure}[]
\centering
\includegraphics[width=.5\textwidth]{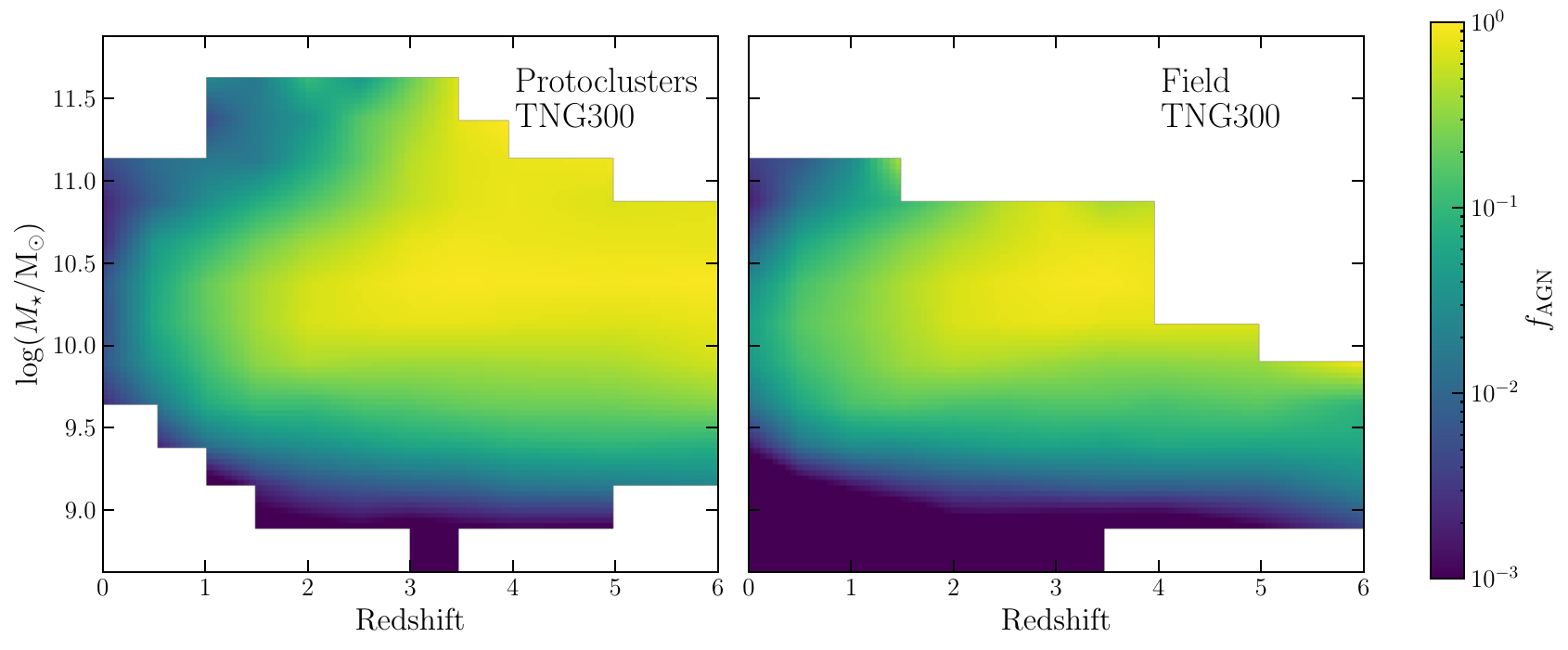}
 \vspace{-0.5cm}
 \caption{Protoclusters (left) and field (right) 2D distributions of the AGN fraction as a function of the galaxy stellar mass and redshift. The map is color coded with the AGN fraction values. Here we consider AGN as those with $L_{\rm BOL} > 10^{44} \, {\rm erg \, s^{-1}}$.}
 \label{fig:2D_map_AGN_frac}
\end{figure}

\par A combined view of the two approaches can be seen in Figure \ref{fig:2D_map_AGN_frac}, where we compare protoclusters and field AGN fraction distributions as a function of both redshift and galaxy stellar mass. These trends indicate that the apparent environmental enhancement of the AGN fraction arises from the different stellar mass distributions of galaxies in the two environments. At fixed stellar mass, the AGN fraction is consistent between field and protoclusters, while differences emerge when integrating over evolving galaxy populations (see Simpson paradox). This suggests that environment primarily influences AGN activity indirectly, by accelerating the assembly of massive galaxies and SMBHs in overdense regions, rather than by directly enhancing SMBH accretion at fixed host galaxy stellar mass, at least at a population-wide level.

\subsection{Bolometric AGN luminosity functions and BH accretion rate density}
While the AGN fraction provides a useful measure of the incidence of active SMBHs above a given luminosity threshold, it is intrinsically conditional on host galaxy properties, most notably stellar mass. As discussed in the previous Sections, this can lead to apparently conflicting trends when comparing different environments, due to the evolving stellar mass distributions of galaxies in the field and in protoclusters.

For this reason, we complement the AGN fraction analysis with a study of the AGN bolometric luminosity function (LF) in the two environments. The LF directly probes the distribution of SMBH accretion luminosities, independently of any explicit selection on host-galaxy mass, and therefore provides a more global and physically motivated characterization of AGN activity across environments.

\subsubsection{AGN LF}\label{subsec:LF}
\begin{figure*}[]
\centering
\includegraphics[width=1.\textwidth]{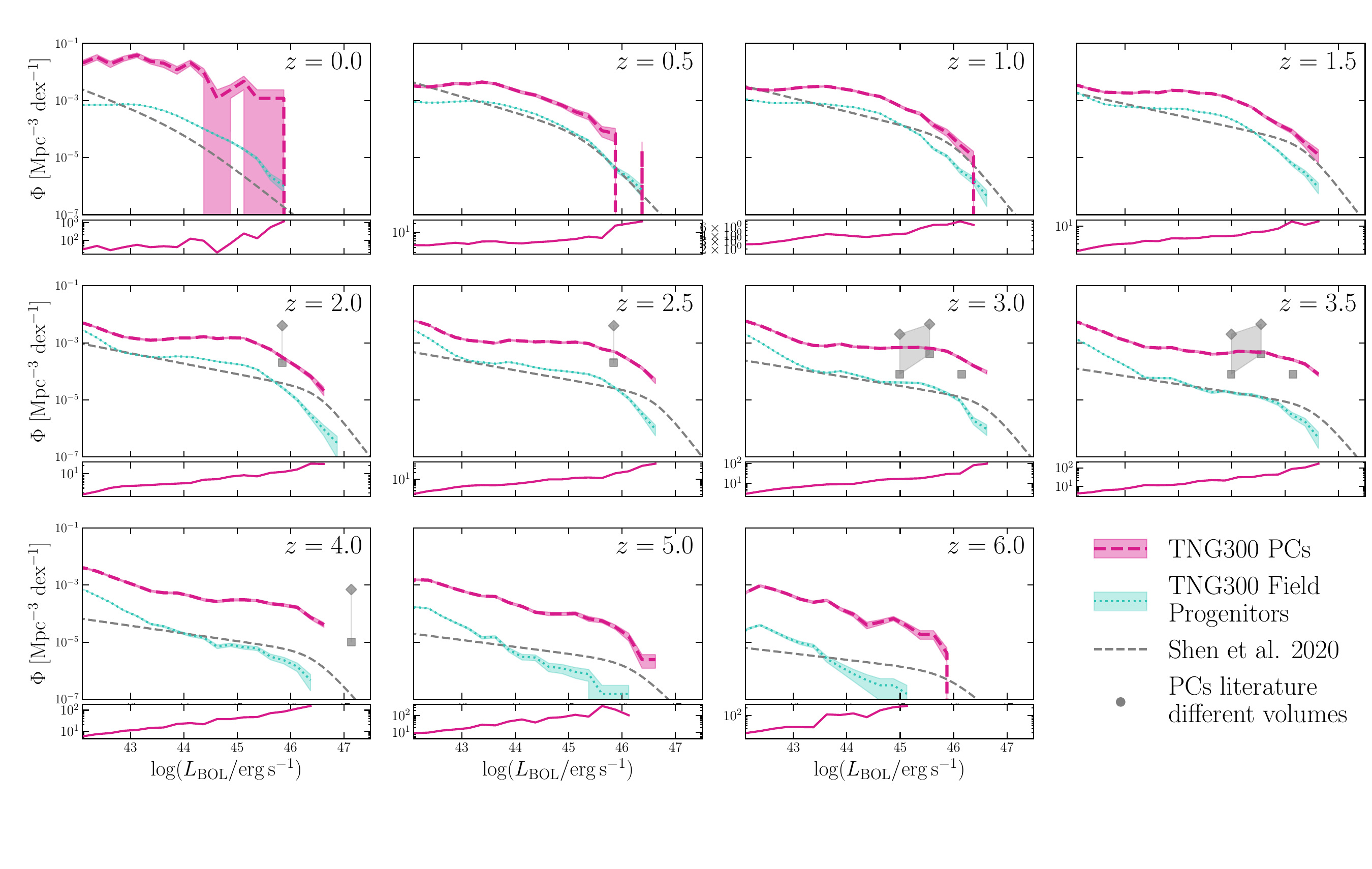}
\vspace{-1cm}
 \caption{AGN bolometric luminosity functions in field (green) and protocluster (red), in different redshift bins. The lower panels are showing the protocluster-to-field ratios. The grey dashed curve is the AGN bolometric LF by \citet{shen2020}, while the grey points are from observed protoclusters, obtained assuming two different volume definition \citep[Slug, Fabulous, J0819, SPT and DRC, see][]{vito2020, vito2024, traina2025}.}
 \label{fig:AGN_LF}
\end{figure*}

To compute the AGN bolometric luminosity function (LF), we evaluate the number density of AGN as a function of bolometric luminosity and redshift, defined as
\begin{equation}
\Phi(L_{\rm BOL}, z) =
\frac{N_{\rm AGN}(L_{\rm BOL})}{\Delta \log L_{\rm BOL}\; V_{\rm env}} \, ,
\end{equation}
where \(N_{\rm AGN}(L_{\rm BOL})\) is the number of AGN within a given bolometric luminosity bin, \(\Delta \log L_{\rm BOL}\) is the bin width, and \(V_{\rm env}\) is the co-moving volume of the considered environment (protoclusters or field).

To define the volume of protoclusters, we adopt a radius definition following \citet{baxter2025}, in which the protocluster radius is defined as the radius enclosing 90\% of the total stellar mass of the descendant cluster at each redshift, \(cR_{90}\). This definition allows the protocluster volume to evolve self-consistently with redshift, capturing the region that is physically associated with the assembly of the descendant cluster. The co-moving volume of each protocluster is then computed assuming a spherical geometry with radius equal to \(cR_{90}\). The total protocluster volume at a given redshift is obtained by summing the individual volumes of all protoclusters in the sample at that redshift. The resulting LF should be interpreted as the average AGN number density within protocluster regions. Although the assumption of spherical symmetry for protocluster volumes is reasonable and has been widely adopted in the literature for these systems \citep[see e.g.,][]{muldrew2015,chaing2017,baxter2025}, it is important to acknowledge that protoclusters are not necessarily spherical structures. In fact, they are often characterized by complex, filamentary morphologies arising from the anisotropic nature of large-scale structure formation. Alternative approaches aimed at better tracing the actual galaxy distribution within protoclusters include the adoption of ellipsoidal geometries, which account for their intrinsic elongation, or the use of convex hull techniques, which provide a more direct, morphology-independent estimate of the occupied volume.

The co-moving field volume is defined as the total co-moving volume of the TNG300 simulation box, excluding the regions associated with the FoFs removed throughout the analysis. Over all the redshift bins considered in this work, the field environment covers most of the simulation volume ($\sim 100\%$ at low redshifts, down to the $\sim 90\%$ at $z = 6$).
By construction, the field volume samples the low-density regions of the simulation, ensuring a clean separation between overdense and average-density environments.

The resulting AGN bolometric luminosity functions (LFs) are shown in Figure \ref{fig:AGN_LF}. Both the field and protocluster LFs are broadly consistent with estimates from the literature, although in some cases the highest luminosities are not fully sampled due to the limited volume of the simulation. As expected, the AGN LFs in protoclusters are systematically higher than those in the field, with protocluster-to-field ratios ranging from \(\sim 10\) up to a few hundreds, depending on redshift and bolometric luminosity. 

The enhancement of AGN activity in protoclusters increases with both redshift and luminosity, being particularly pronounced at the high-luminosity end, which reflects the early growth of the most massive SMBHs in overdense environments, as shown in the lower panels of Figure \ref{fig:AGN_LF}, where the ratio between the protocluster and field AGN-LF is plotted as a function of bolometric luminosity.
The increasing enhancement toward the bright end of the AGN luminosity function further supports the interpretation emerging from the AGN fraction analysis and it is in agreement with observational findings by \citet{tozzi2022a} and \citet{travascio2025}. Although the AGN fraction at fixed stellar mass is broadly consistent between protoclusters and field galaxies, protoclusters host a larger abundance of massive galaxies and SMBHs, particularly at high redshift. Since more massive SMBHs are also capable of reaching higher bolometric luminosities, differences in the underlying galaxy and SMBH mass distributions naturally propagate into a luminosity-dependent enhancement in the AGN LF. This effect is particularly evident in the lower panels of Figure~\ref{fig:AGN_LF}, where the protocluster-to-field enhancement increases toward high luminosities.

We note that this interpretation may appear in tension with the results presented by \citet{IslaLlave2026}, in which they found an enhanced AGN fraction in protoclusters even after matching the stellar-mass distributions of galaxies. However, their analysis focused specifically on sub-millimeter galaxies (SMGs), i.e. a highly star-forming and gas-rich galaxy population, which may preferentially trace phases of enhanced SMBH fueling and merger-driven accretion and are also difficult to be reproduced in simulations. Our results instead describe the average AGN population across the full galaxy population, suggesting that environmental triggering may be important for specific classes of galaxies, while the global enhancement is primarily driven by demographic differences between protocluster and field populations.

\subsubsection{BHAD evolution with redshift}
\begin{figure*}[]
\centering
\includegraphics[width=1.\textwidth]{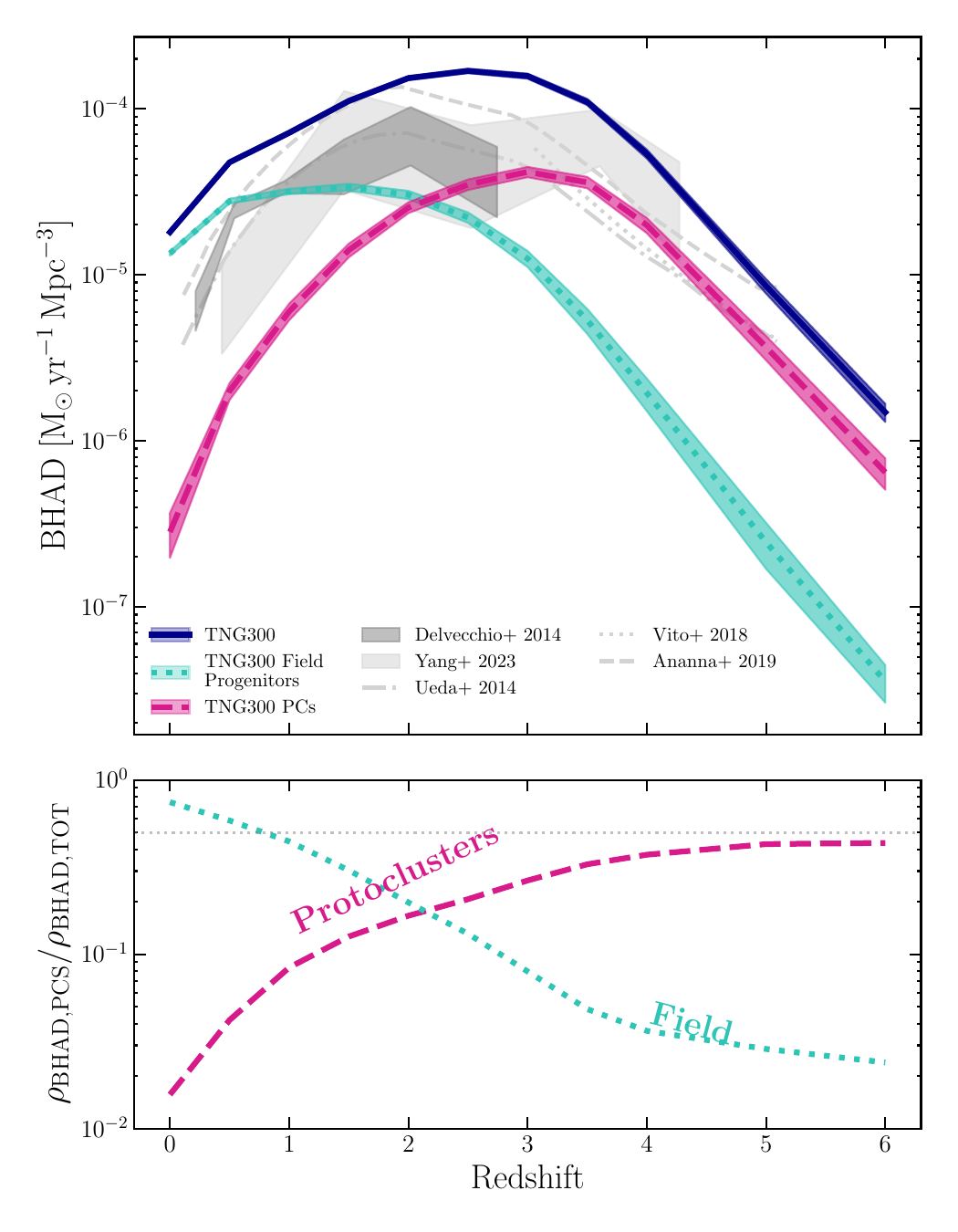}
\vspace{-0.5cm}
 \caption{Redshift evolution of the SMBH accretion rate density in the TNG300 simulation (top panel) and fractional contribution of field and protoclusters (bottom panel). The total, field and protocluster BHAD are shown in blue, green and red, respectively. Errors are computed by integrating the lower and upper curves of the BH accretion rate function. Grey curves and dashed areas are observational results from the literature \citep[][]{delvecchio2014bhard, ueda2014, vito2018, ananna2019, yang2023}.}
 \label{fig:BHAD}
\end{figure*}

Since the accretion rates are available for each SMBH in our sample, we construct SMBH accretion rate functions at each redshift, similarly to what we did for the bolometric AGN LF, and integrate them to study the redshift evolution of the SMBH accretion rate density (BHAD) over the range \(0 \leq z \leq 6\). This approach allows us to disentangle the relative contributions of field and protocluster SMBHs to the total accretion density of the Universe. To derive the accretion rate functions, we adopt a lower bolometric luminosity threshold of \(L_{\rm BOL} = 10^{43}\,{\rm erg\,s^{-1}}\), which roughly corresponds to the observational limit commonly used to define AGN samples.

Figure \ref{fig:BHAD} shows the total BHAD predicted by the simulation, together with the separate contributions from field and protocluster environments, and a compilation of observational estimates from the literature. The total BHAD is broadly consistent with the upper envelope of the observational measurements, which are on average slightly lower in normalization.

The lower panel of Figure \ref{fig:BHAD} highlights the fractional contribution of the different environments to the total BHAD. At low redshift (\(z < 1\)), the field dominates the accretion density, while its contribution rapidly decreases toward higher redshifts. In contrast, protoclusters contribute only a minor fraction of the BHAD in the local Universe, but become increasingly important at earlier cosmic times, dominating the accretion budget at \(z \gtrsim 2\,-\,3\) and contributing up to \(\sim 50\%\) of the total BHAD at \(z \sim 6\). We note that this analysis does not include the contribution from intermediate-density environments, such as galaxy groups and cluster outskirts, which are expected to provide a non-negligible fraction of the total BHAD.

\section{Conclusions}\label{sec:conclusion}

In this work, we investigated the impact of environment on the growth and activity of SMBHs in galaxy protoclusters (i.e., galaxies which are progenitors of $z = 0$ galaxies within $R_{\rm 200, c}$ of clusters FoF halos) from $z = 6$ to the present day, using the TNG300 cosmological hydrodynamical simulation of galaxies of the IllustrisTNG suite. By consistently comparing protocluster and field populations, we explored galaxy and SMBH demography, AGN fractions, bolometric luminosity functions, and the black-hole accretion rate density. Our main results can be summarized as follows.

\begin{itemize}
    \item Protoclusters host systematically more massive galaxies and SMBHs than the field at all redshifts, with the difference becoming more pronounced toward low redshift. In addition, signatures of accelerated evolution are predicted by TNG300 in overdense environments, where galaxies exhibit suppressed star formation at earlier epochs compared to their field counterparts (see Figures 2 to 5).
    \item The AGN fraction increases with stellar mass in both environments (Figure \ref{fig:fagn_zbin}) and shows a similar overall behavior in protoclusters and field when controlling for host-galaxy stellar mass. This indicates that, at fixed galaxy stellar mass, the probability of hosting an AGN is comparable in the two environments.
    \item When analyzed as a function of redshift, the AGN fraction appears enhanced in protoclusters, particularly at high redshift and low stellar masses (see Figure \ref{fig:fagn_mbin}). However, this apparent discrepancy is reconciled when accounting for the different stellar-mass distributions of galaxies in the two environments, highlighting the importance of selection effects when interpreting environmental trends.
    \item The AGN bolometric luminosity function is significantly enhanced in protoclusters compared to the field, with differences increasing toward higher redshift and higher luminosities (top and bottom panels in Figure \ref{fig:AGN_LF}). This reflects the overdense nature of these regions, where massive galaxies and actively accreting SMBHs are more abundant.
    \item Finally, protoclusters contribute an increasingly large fraction of the total BHAD toward high redshift, reaching up to $\sim 50\%$ at $z \sim 6$, despite occupying a small fraction of the cosmic volume (see bottom panel of Figure \ref{fig:BHAD}). This result suggests that overdense regions play a key role in driving the early growth of SMBHs.
\end{itemize}
\par Overall, our results provide a coherent interpretation of the reported enhancement of AGN activity in protoclusters, showing that it primarily arises from the underlying galaxy population rather than from a direct environmental triggering of SMBH accretion. This work highlights the importance of accounting for host-galaxy properties when investigating environmental effects, and provides a robust framework for interpreting current and future observations of AGN in dense environments. Future works may involve different overdensity selection criteria and other simulations in order to improve statistics, get even a more robust characterization of the AGN population in overdense environments, and alleviate potential systematics present in simulations.

\section*{Acknowledgements}
AT, FV, CV, RG, MNIL, VC acknowledge support from the “INAF Ricerca Fondamentale 2023 \& 2024 - Large GO” grant. AP, AK acknowledge funding
from the European Union (ERC, COSMIC-KEY, 101087822, PI: Pillepich).

\bibliographystyle{aa}
\bibliography{1biblio}

\begin{appendix}

\section{Halo mass function: protoclusters vs field}\label{app:halo_MF}
We derive the halo mass function (HMF) for haloes in protoclusters and field. At each redshift, we compute the HMF using the volume defined as described in Section \ref{subsec:LF} for the AGN LF. At high redshifts, the protoclusters and field HMFs are almost complementary, with the first covering the higher masses and the second dominating the lower masses. At $z = 0.5$ the normalization of the two HMFs is similar, but the mass range is still different. Finally, at $z = 0$, the volume density of protocluster haloes is much higher than the volume density of field haloes.
\par The large mass range probed by the protocluster haloes is the main cause of the larger stellar mass range covered by protocluster galaxies.

\begin{figure*}[]
\centering
\includegraphics[width=1.\textwidth]{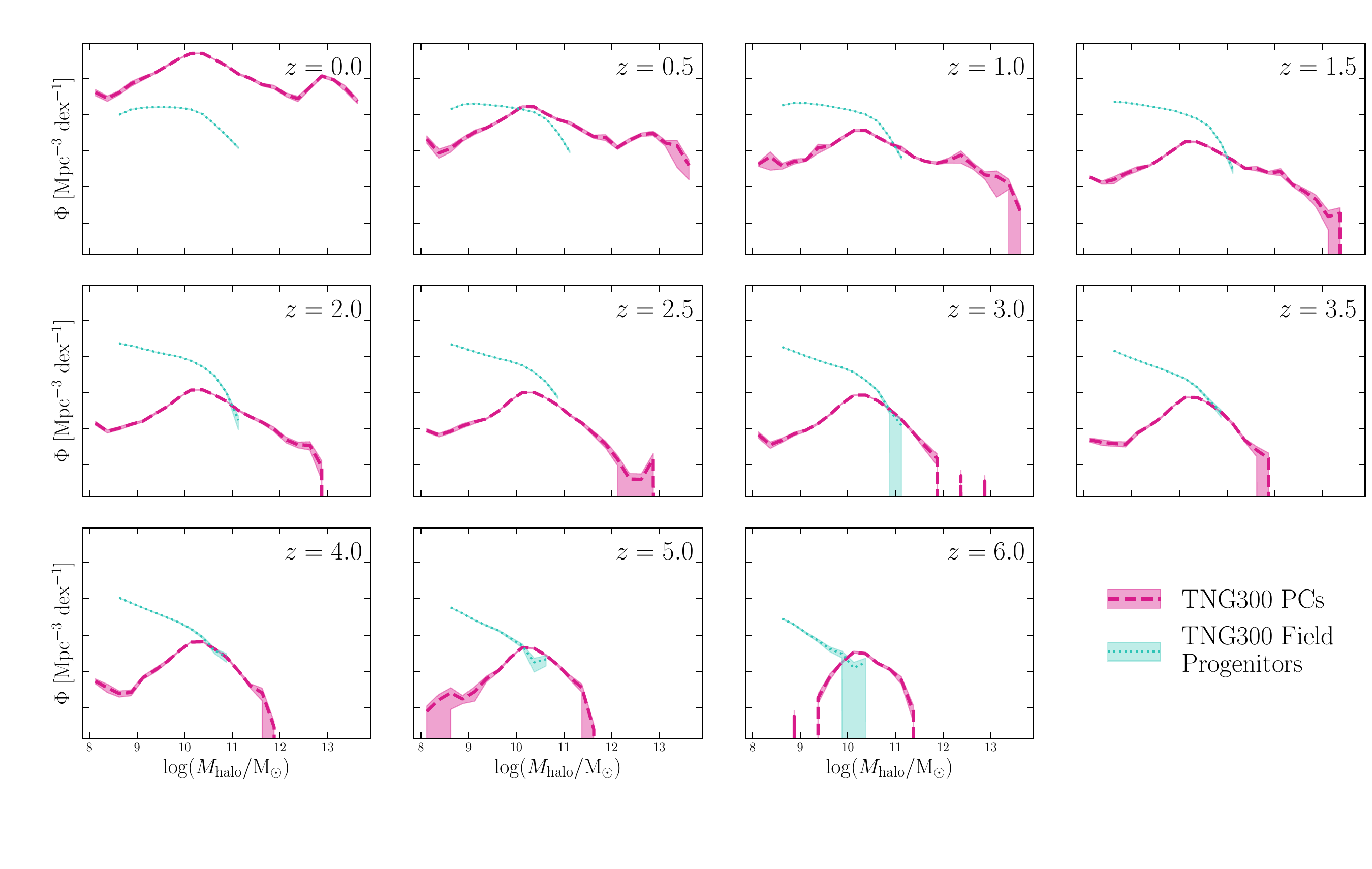}
 \caption{Halo mass function at the different redshift bins, for protoclusters (red) and field (green) galaxies.}
 \label{fig:halo_mf}
\end{figure*}

\section{Gas content: protoclusters vs field}\label{app:gas_content}

\begin{figure*}[]
\centering
\includegraphics[width=1.\textwidth]{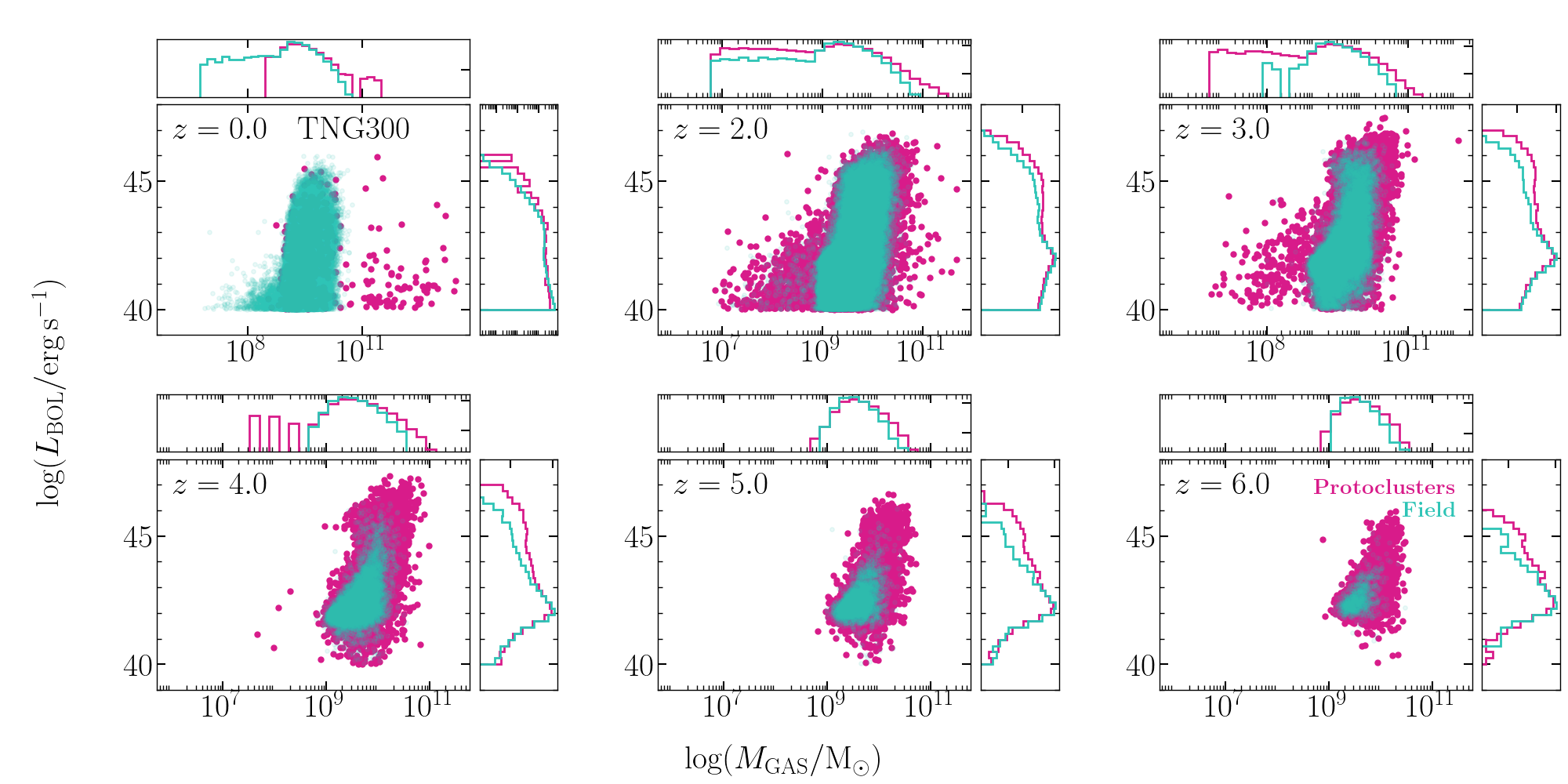}
 \caption{Same as Figure \ref{fig:SFR_Mstar} but for $L_{\rm BOL} \,-\, M_{\rm GAS}$, i.e., the distributions of SMBHs bolometric luminosity and galaxy gas mass.}
 \label{fig:Lbol_Mgas}
\end{figure*}

In this appendix, we show the distribution of the AGN bolometric luminosity as a function of their gas content (see Figure \ref{fig:Lbol_Mgas}). At high redshifts, more luminous AGN are hosted by galaxies richer in gas content: bright AGN are powered by large gas reservoirs available in the galaxies to be consumed during the accretion process. Moreover, protoclusters host more gas rich galaxies then the field, at fixed gas mass. Since $z = 3$ down to $z = 0$, the gas content begins to be suppressed in both environments, as it is consumed to power accretion and star formation. Indeed, at low redshifts, less luminous AGN are progressively poorer in gas mass.

\end{appendix}

\end{document}